\documentclass[pra,twocolumn]{revtex4}
\usepackage{graphicx,amsmath}

\newcommand{\be}{\begin{equation}}
\newcommand{\ee}{\end{equation}}
\newcommand{\bz}{\overline{z}}
\newcommand{\ket}{|z\rangle}
\newcommand{\bra}{\langle z|}
\newcommand{\vars}{(\bz_f,z_i,t)}
\newcommand{\cI}{\mathcal{I}_{SK}}
\newcommand{\cS}{\mathcal{S}}

\begin{document}
\title{Semiclassical Husimi functions for spin systems}
\author{Marcel Novaes}
\email{mnovaes@ifi.unicamp.br}
\author{Marcus A.M. de Aguiar}
\affiliation{Instituto de F\'{i}sica ``Gleb Wataghin",
Universidade Estadual de Campinas, 13083-970 Campinas-SP, Brazil}

\begin{abstract}
We derive a semiclassical approximation to the Husimi functions of
stationary states of spin systems. We rederive the Bohr-Sommerfeld
quantization for spin by locating the poles of the corresponding
local Green function. The residues correspond to the Husimi
functions, which are seen to agree very well with exact
calculations.
\end{abstract}

\maketitle

Rigorous derivation of semiclassical approximations in phase space
via path integrals \cite{prd19jrk1979} for systems with one degree
of freedom have recently received considerable attention, both for
continuous variables and spin systems. Baranger {\it et al}
\cite{jpa34mb2001}, for example, have discussed the canonical
coherent state path integral and its semiclassical approximation
in some detail, including an initial-value representation and the
Green function. The study of semiclassical propagation of wave
packets, using complex \cite{pre47st1993} or nearly real
\cite{pra66tvv2002} trajectories, for regular and chaotic
\cite{pre69adr2004} systems, has developed considerably over the
last few years. The spin path integral, and its semiclassical
approximation, has found an important application in the study of
spin tunnelling and topological effects \cite{prl69dl1992}. Stone
{\it et al} have derived the spin coherent state semiclassical
propagator in detail \cite{jmp41ms2000}, paying particular
attention to the so-called Solari-Kochetov \cite{jmp28hgs1987}
correction. This correction is related to the difference between
the average value of the Hamiltonian in coherent states and its
Weyl symbol \cite{jmp45mp2004}, and has a counterpart in the
canonical case \cite{jpa34mb2001}.

To obtain semiclassical approximations for the energy levels $E_n$
and stationary states $\langle x|n\rangle=\psi_n(x)$ of
one-dimensional bound systems, on the other hand, one normally
resorts to the usual Bohr-Sommerfeld (BS) and WKB theories
\cite{brack1997}. A coherent state version of these theories,
which works in phase space, is also available \cite{jpa34mb2001}
and produces a BS formula and a semiclassical approximation to the
Husimi functions $\mathcal{H}_{n}(z)=|\langle z|n\rangle|^2$.
Recently, Garg and Stone \cite{prl92ag2004} have derived a
semiclassical (BS-like) quantization condition for spin systems,
including the first quantum corrections (see also
\cite{pra40jk1989}). By taking the trace of the semiclassical
Green function, they obtained the energy levels as the location of
its poles. In the present work we have obtained the semiclassical
Husimi functions for spin systems.

The non-normalized spin coherent states are defined by
$\ket=\exp\{zJ_+\}|j,-j\rangle$, and the semiclassical
approximation to the propagator $K=\langle
z_f|e^{-i\hat{H}t/\hbar}|z_i\rangle$ is \cite{jmp41ms2000} \be
\label{prop}
K_{scl}\vars=\left(\frac{i}{\hbar}\frac{e^{i\mathcal{B}/
j}}{2j}\frac{\partial^2 S}{\partial z_i\partial
\bz_f}\right)^{1/2}\exp \left\{\frac{i}{\hbar}\Phi\right\},\ee
where the phase is the classical action plus an extra term known
as the Solari-Kochetov (SK) correction: \be \Phi=
S+\cI=S+\int_0^tA(t')dt'.\ee The classical spin action is given by
\be S=\int_0^t \left[i\hbar j\frac{\bz \dot{z}-\dot{\bz}z}{1+\bz
z}-H(z,\bz)\right]dt'+\mathcal{B}, \ee where the integral is done
along the classical trajectory determined by the Hamilton
equations of motion \be
-i\hbar\dot{\bz}=\frac{1}{g(z,\bz)}\frac{\partial H}{\partial z},
\quad i\hbar\dot{z}=\frac{1}{g(z,\bz)}\frac{\partial H}{\partial
\bz},\ee and the classical Hamiltonian $H(z,\bz)$ is the average
value of the quantum Hamiltonian, $H(z,\bz)=\bra \hat{H}\ket/\bra
z\rangle$. This action obeys the Hamilton-Jacobi relations \be
\frac{i}{\hbar}\frac{\partial S}{\partial
\bz_f}=\frac{2jz(t)}{1+\bz_f z(t)}, \quad
\frac{i}{\hbar}\frac{\partial S}{\partial
z_i}=\frac{2j\bz(0)}{1+\bz(0)z_i}, \quad \frac{\partial
S}{\partial t}=-H.\ee The function $g(z,\bz)$, which is unity in
the canonical case, is given by \be g(z,\bz)=\frac{\partial
^2}{\partial z
\partial \bz}\ln\bra z\rangle=\frac{2j}{(1+z\bz)^2}.\ee

Note that $\mathcal{B}=-i\hbar j \ln[(1+\bz_fz(t))(1+\bz(0)z_i)]$
is a boundary term, that takes into account the fact that in
general $\bz$ is not the complex conjugate of $z$ (the discrete
time formulation of the path integral indicates that the variables
$z$ and $\bz$ must be considered as independent, so we denote the
actual complex conjugate of $z$ by $z^\ast$). That means that if
one defines the usual canonical $(q,p)$ variables according to \be
\label{cano} \frac{z}{\sqrt{1+z\bz}}=\frac{q+ip}{\sqrt{4\hbar j}},
\quad \frac{\bz}{\sqrt{1+z\bz}}=\frac{q-ip}{\sqrt{4\hbar j}} ,\ee
then $q$ and $p$ will in general be complex numbers.

The semiclassical limit for spin systems consists in letting
$\hbar\to 0$ and $j\to \infty$, but keeping $\hbar j=1$. If the
Hamiltonian if $O(\hbar j)$, then $S$ is $O(\hbar j)$, but the SK
correction \be A=\frac{\partial}{\partial
\bz}\frac{1}{4g(z,\bz)}\frac{\partial H}{\partial
z}+\frac{\partial}{\partial z}\frac{1}{4g(z,\bz)}\frac{\partial
H}{\partial \bz}\ee is $O(\hbar)$, and therefore can be considered
small. Note that since $\bz\neq z^\ast$ the Hamiltonian
$H(z,\bz)$, the action and the SK correction can all be complex.

The semiclassical Green function, \be \label{green}
G_{scl}(z,E)=\frac{1}{i\hbar}\int_0^\infty K_{scl}(\bz,z,t)
e^{iEt/\hbar}dt,\ee can be calculated by making a stationary
exponent approximation to the integral. Note that we are
interested only in its diagonal elements. This implies $z_i=z$,
$\bz_f=z^\ast$, but in general $z(t)\neq z$ and $\bz(0)\neq
z^\ast$, so that we do not have a real periodic orbit (by real
orbit we mean one in which $q$ and $p$ are real). The stationary
time $t_0$ is determined by the condition \be \label{stat}
\left.\frac{d(\Phi+Et)}{dt}\right|_{t_0}=\left.\frac{\partial
S}{\partial t}\right|_{t_0}+\left.\frac{\partial \cI}{\partial
t}\right|_{t_0}+E=0.\ee As usual in semiclassical calculations, we
do not consider derivatives of $A$, because including such terms
would be inconsistent with the gaussian approximation involved in
the derivation of (\ref{prop}). Therefore Eq.(\ref{stat}) can be
also written as \be E-\mathcal{E}(z,t_0)+A(z,t_0)=0,\ee where \be
\mathcal{E}(z,t_0)=-\left.\frac{\partial S}{\partial
t}\right|_{t_0}\ee is the energy of the classical trajectory, not
to be confused with $E$, the argument of the Green function. In
order to proceed with the integration, we need to expand the
exponent to second order in time. We define \be
\left.\frac{\partial ^2 S}{\partial
t^2}\right|_{t_0}:=\alpha(z,t_0),\ee and neglect the second
derivative of $I_{SK}$, in order to obtain \be
G(z,E)=\frac{1}{i\hbar}\left(-\frac{\pi e^{i\mathcal{B}/\hbar
j}}{\alpha j}\frac{\partial^2 S}{\partial z_i\partial
\bz_f}\right)_{t_0}^{1/2}\exp\{\frac{i}{\hbar}\varphi\},\ee where
\be\varphi=S(t_0)+\cI(t_0)+Et_0.\ee We can find a more convenient
way of expressing $\alpha$ in order to transform the prefactor.
The form (see \cite{jpa34mb2001}) \be \alpha=
-\dot{z}\dot{\bz}\frac{\partial ^2 S}{\partial \bz_f\partial
z_i}\ee leads to \be G(z,E)=\frac{1}{i\hbar}\left(\frac{\pi
e^{i\mathcal{B}/\hbar j}}{\dot{z}\dot{\bz}
j}\right)^{1/2}\exp\{\frac{i}{\hbar}\varphi\}.\ee

Even with this simplification it is hard to find the poles of
$G(z,E)$. Garg and Stone \cite{prl92ag2004} have done this by
calculating its trace under another stationary phase
approximation, which leads to $z(t)=z$, $\bz(0)=z^\ast$, and thus
to real periodic orbits. We take a different route, that will
allow us to obtain not only the energy levels but also the Husimi
distributions. Even though the classical orbits involved in the
calculation of (\ref{prop}) and (\ref{green}) are complex, we
argue that the largest contributions to the function $G(z,E)$ (and
not only to its trace) must come from the vicinity of the real
periodic orbit through $z$, and its repetitions. The accuracy of
the final results support this idea.

\begin{figure}[b]
\includegraphics[scale=0.25,angle=-90]{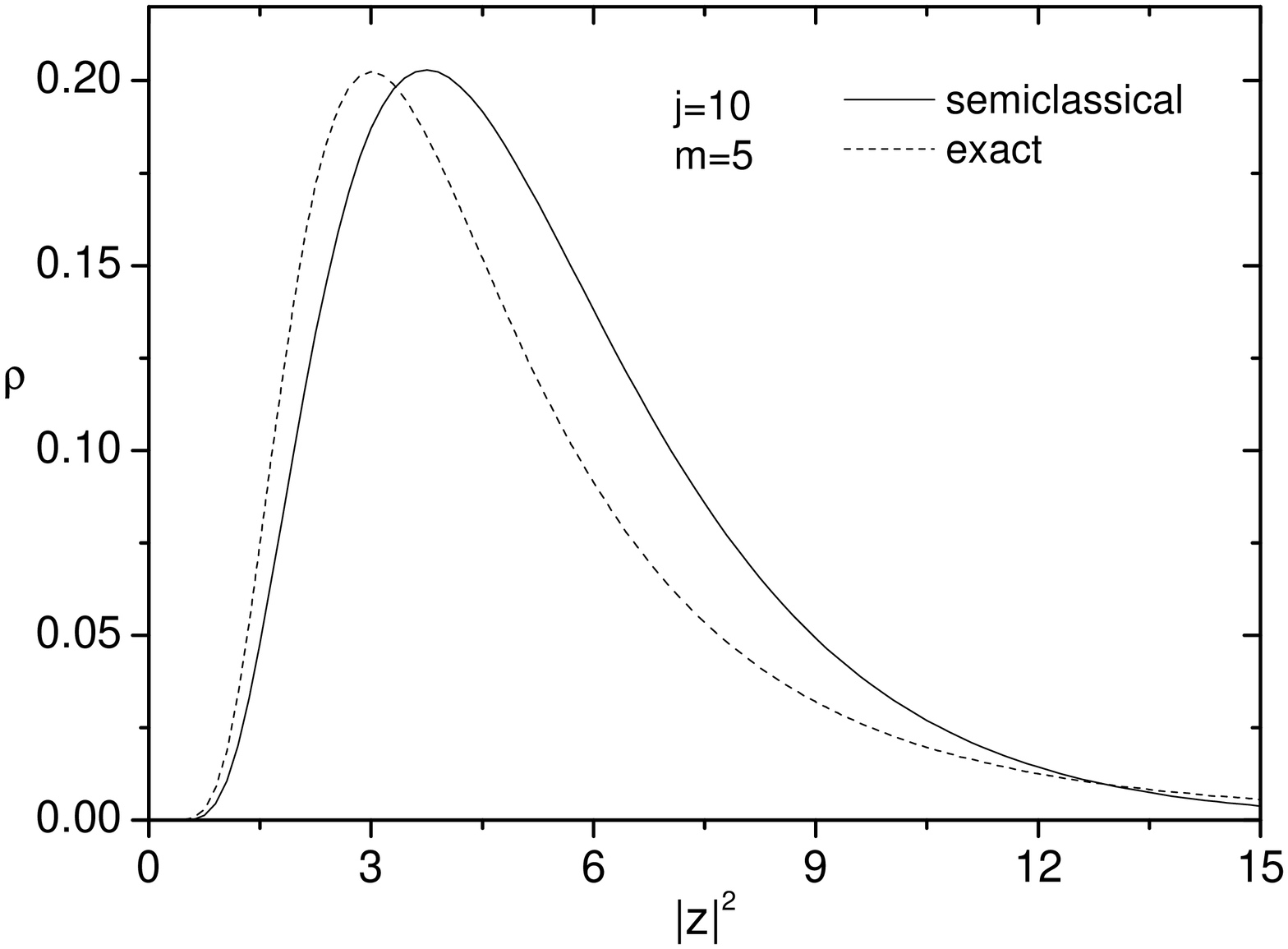}\\
\includegraphics[scale=0.25,angle=-90]{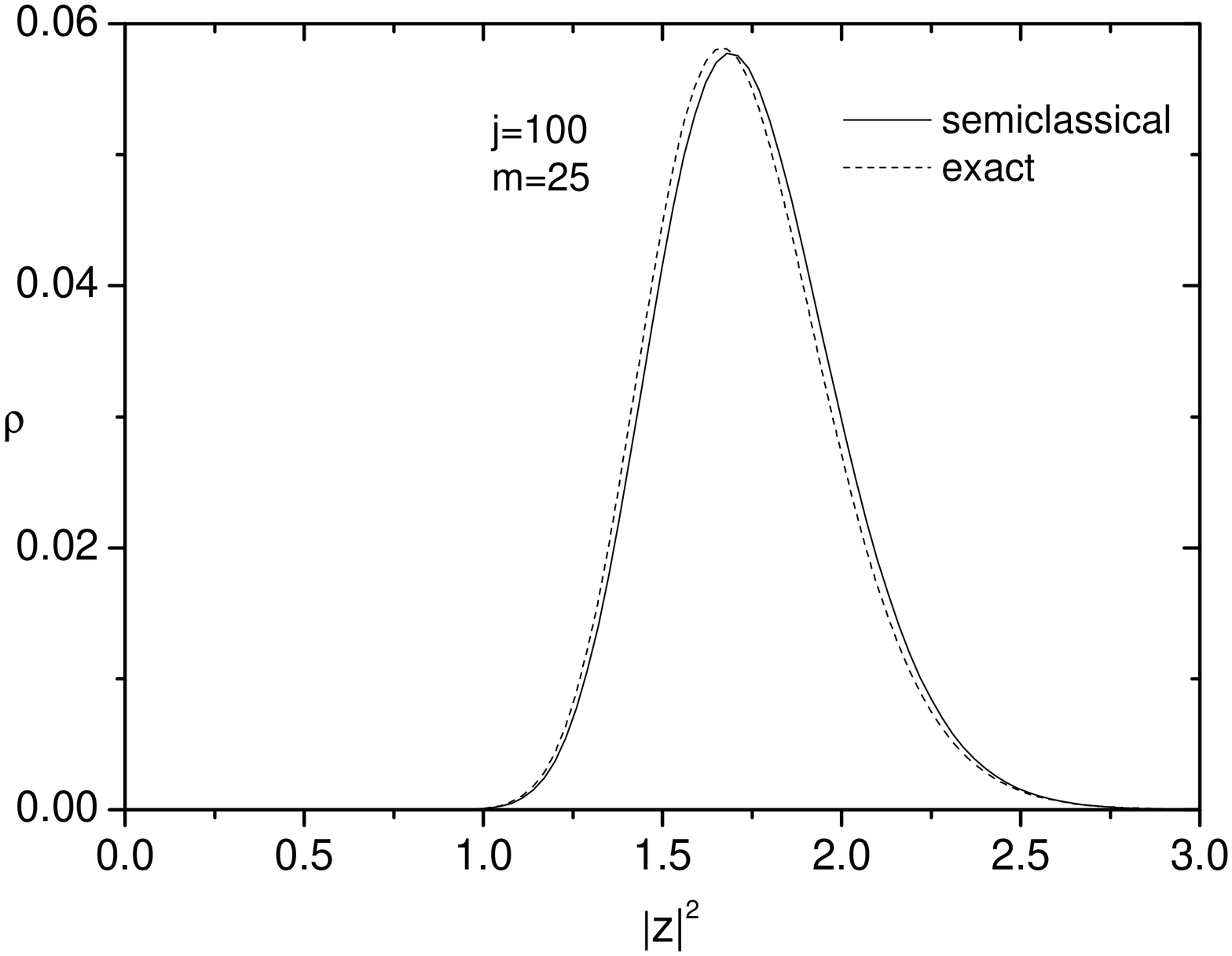}
\caption{Exact (top) and semiclassical (bottom) Husimi
distributions for the simple Hamiltonian $H=\omega\hbar J_z$.}
\end{figure}

Let us denote the period of the orbit through $z$ by $T(z)$ and
expand the stationary time $t_0$ as \be t_0\simeq nT+T_0,\ee where
$n$ counts the repetitions of the real periodic orbit and $T_0$,
assumed small, has to be determined. Expanding the stationary
exponent condition (\ref{stat}) we find \be
T_0=-\frac{E-\mathcal{E}+A}{\alpha^{(n)}},\ee where \be
\alpha^{(n)}:=\alpha(z,nT)=-\left.\frac{\partial \mathcal{E}
}{\partial t}\right|_{nT}=\left.\frac{\partial ^2 S}{\partial
t^2}\right|_{nT}.\ee

\begin{figure}[b]
\includegraphics[scale=0.25,angle=-90]{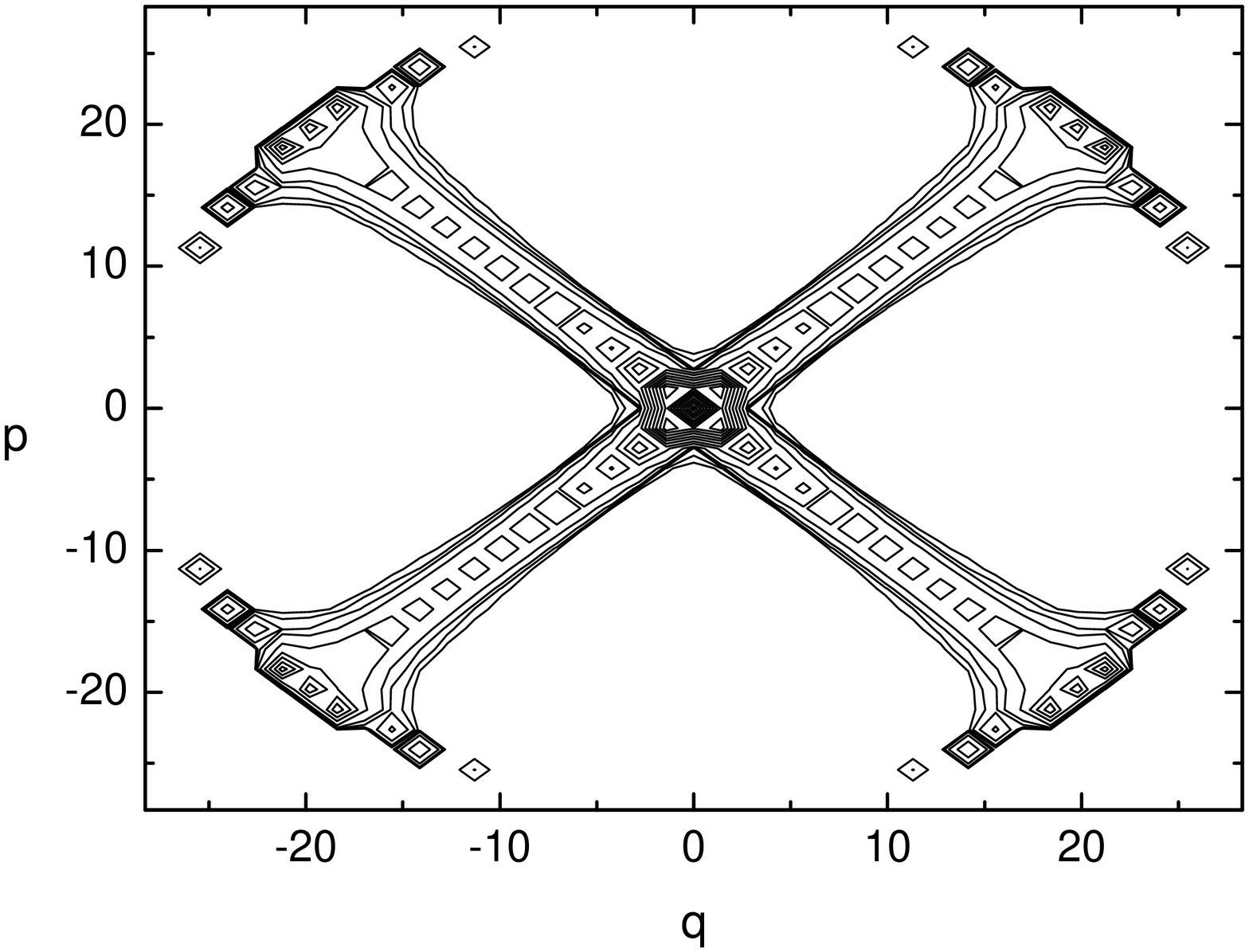}\\
\includegraphics[scale=0.25,angle=-90]{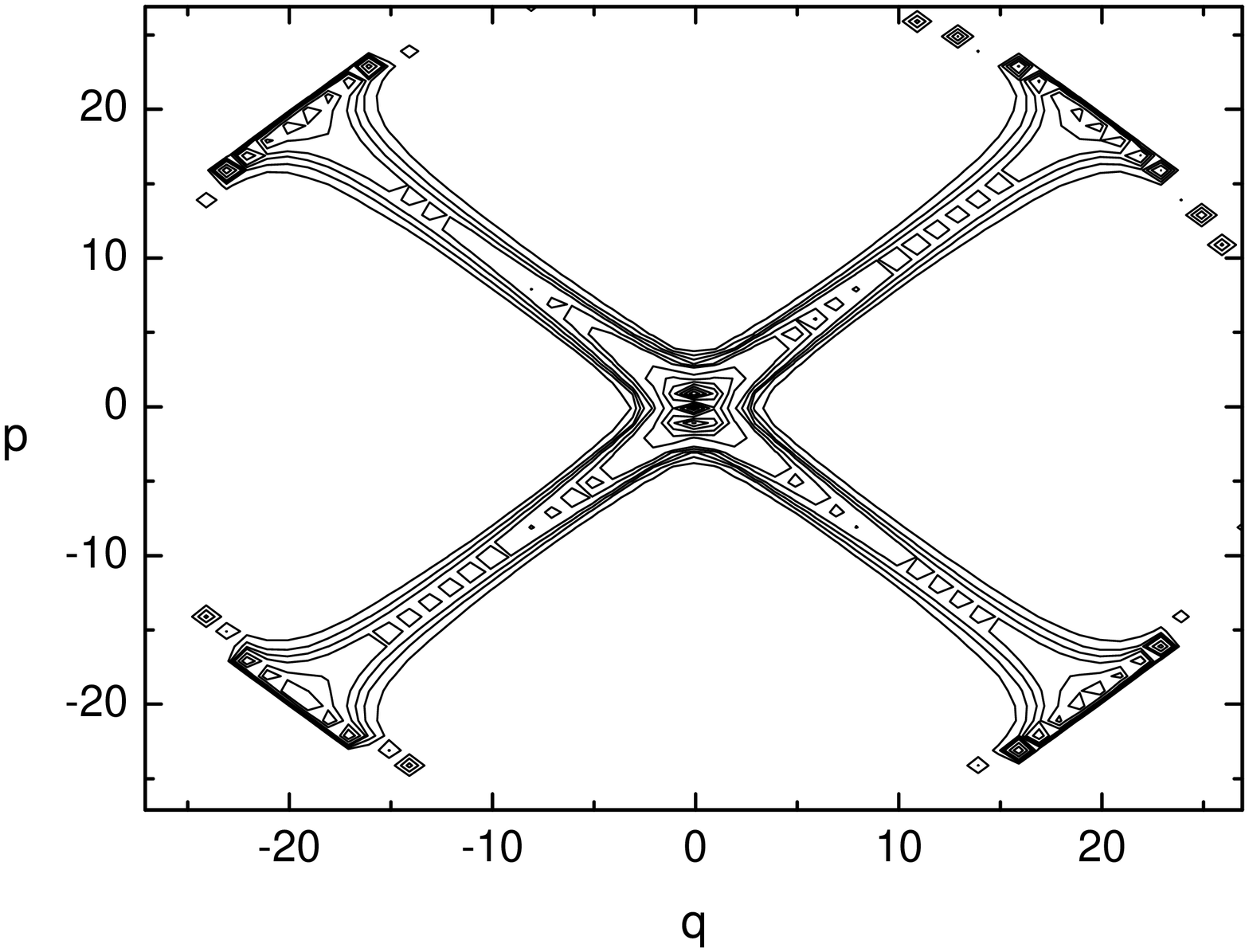}
\caption{Exact (top) and semiclassical (bottom) Husimi
distributions for the $200{\rm th}$ state of the LMG model, with
$j=200$, $\omega=1$ and $\hbar\alpha=1000$.}
\end{figure}

Now we must expand $\varphi$ to second order in $T_0$. Note that
$S(nT)=n\cS-n\mathcal{E}T+\mathcal{B}(nT)$, where \be \cS=i\hbar
j\int_0^T\frac{\bz \dot{z}-\dot{\bz}z}{1+\bz z}dt.\ee Note also
that $\cI(nT)=n\cI(T)$. It can also be shown \cite{jpa34mb2001}
that \be \label{alpha} \frac{1}{\alpha^{(n)}}=\frac{1}{i\hbar
g(z,\bz) |\dot{z}|^2}-n\frac{d^2 \cS}{d \mathcal{E}^2}.\ee After
$n$ repetitions of a periodic orbit the prefactor acquires a phase
of $(-1)^n$. Therefore, the result of this expansion is
\begin{align} \varphi & \simeq
n\left[\cS-\mathcal{E}T+ET+\cI-\pi\hbar\right]\nonumber\\&+\mathcal{B}(nT)-\mathcal{E}T_0+
\frac{\alpha^{(n)}}{2}T_0^2+ET_0+AT_0.\end{align} If we add and
subtract $nAT$, define $x=E-\mathcal{E}+A$ and use
$T=d\cS/d\mathcal{E}$ together with equation (\ref{alpha}) we
obtain
\begin{align} \varphi & \simeq
n\left[\cS+\cI-\pi\hbar+\frac{d\cS}{d\mathcal{E}}x-
\frac{d\cS}{d\mathcal{E}}A+\frac{1}{2}\frac{d^2 \cS}{d
\mathcal{E}^2}x^2\right]\nonumber\\&-\frac{1}{g(z,\bz)}\frac{x^2}{2i\hbar
|\dot{z}|^2}+\mathcal{B}(nT).
\end{align}

\begin{figure}[b]
\includegraphics[scale=0.25,angle=-90]{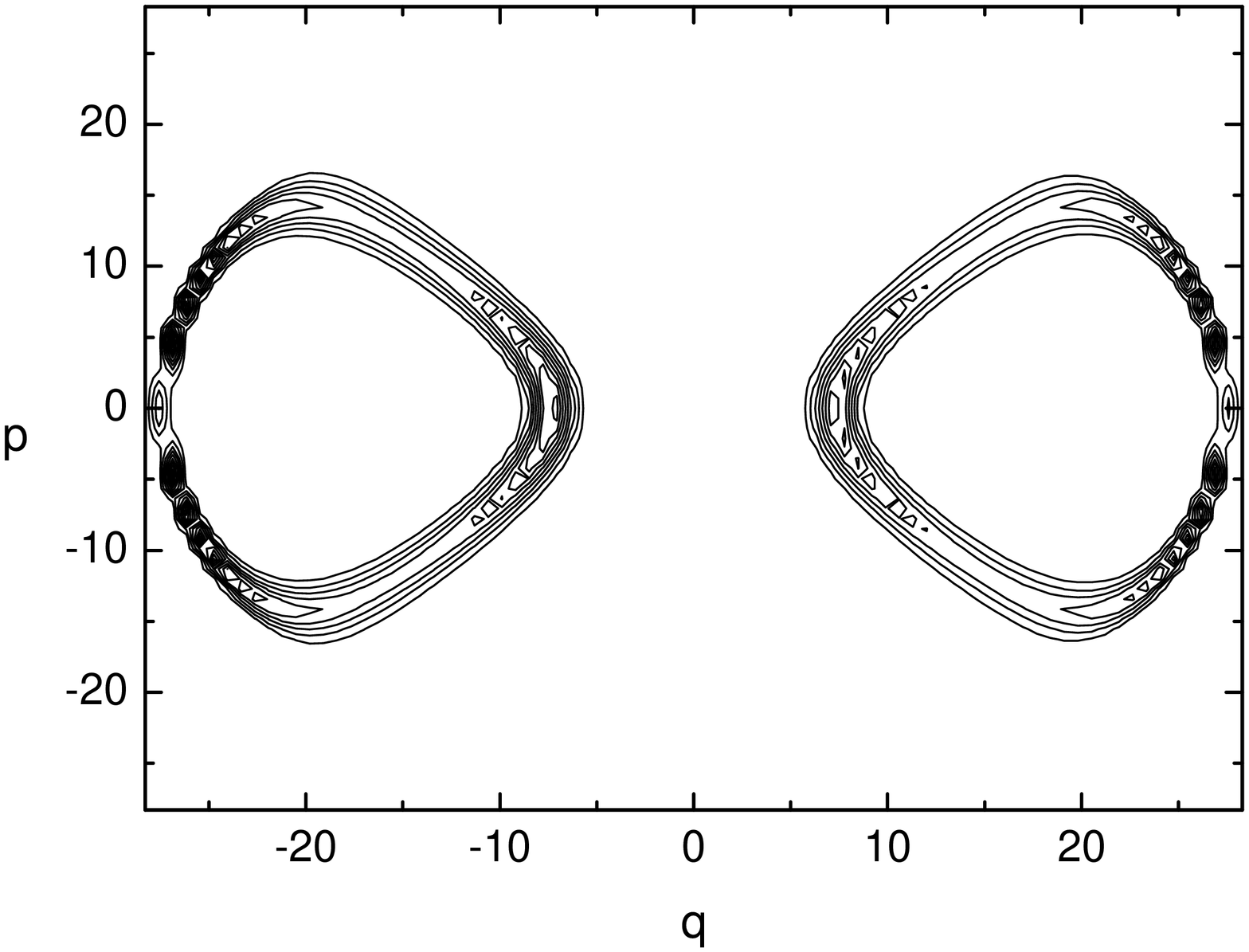}\\
\includegraphics[scale=0.25,angle=-90]{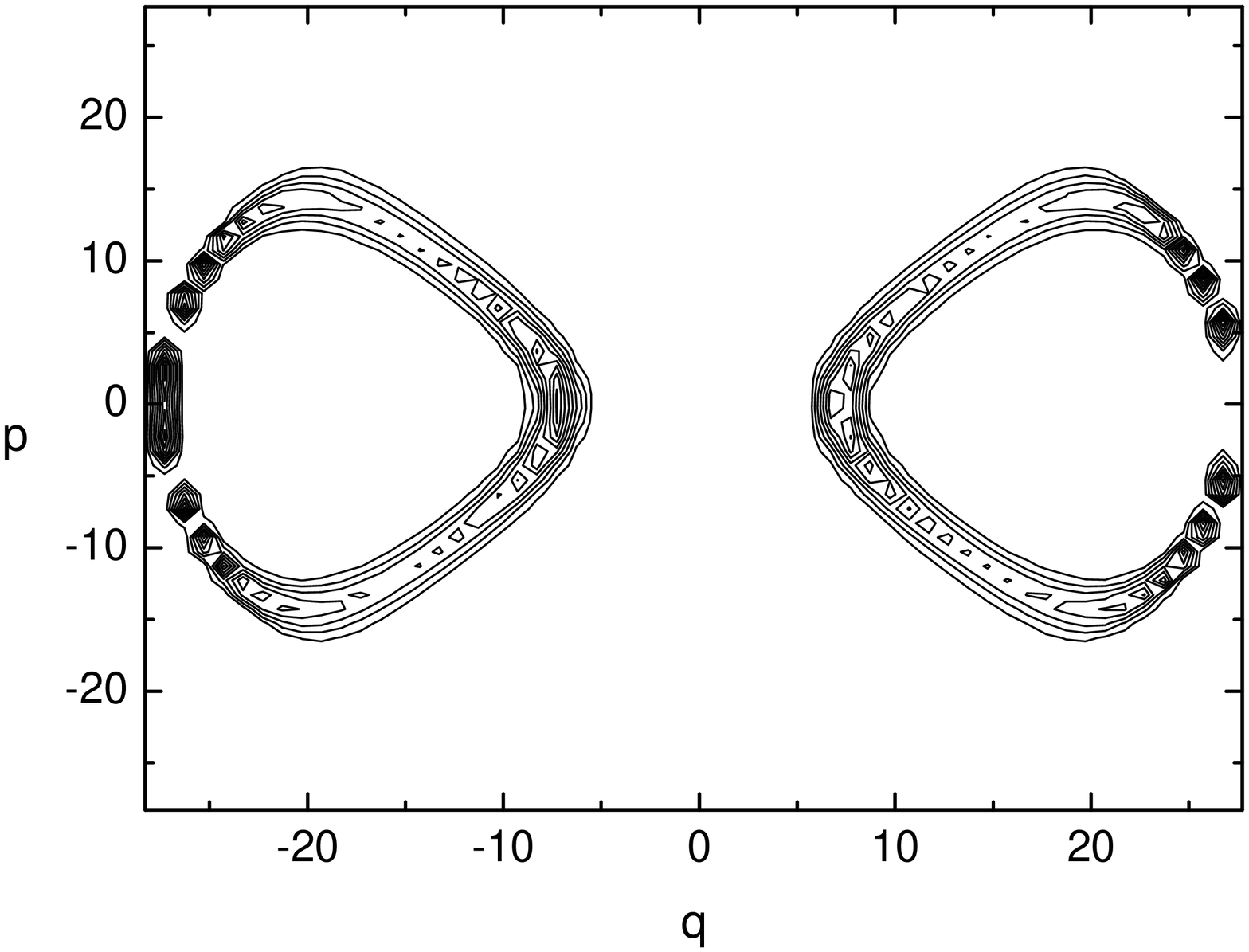}
\caption{Exact (top) and semiclassical (bottom) Husimi
distributions for the $270{\rm th}$ state of the LMG model. The
parameters are the same as in the previous figure. }
\end{figure}

We recognize inside the brackets the expression for the expansion
of $\cS(E+A)$ around $\mathcal{E}$. Since $A\sim O(\hbar)$ we
further expand \be \cS(E+A)\simeq
\cS(E)+\frac{d\cS}{d\mathcal{E}}A\ee and we end up with \be
\varphi \simeq
n[\cS(E)+\cI(E)-\pi\hbar]-\frac{1}{g(z,\bz)}\frac{x^2}{2i\hbar
|\dot{z}|^2}+\mathcal{B}(nT).\ee Summing over $n$ we obtain
\begin{align} G(z,E)&=\frac{\sqrt{\pi}(1+|z|^2)^{2j+1}}{i\hbar
j|\dot{z}|}\frac{e^{i[\cS+\cI-\pi\hbar]/\hbar}}{1-e^{i[\cS+\cI-\pi\hbar]/\hbar}}\nonumber\\
&\times\exp\left\{-\frac{1}{g(z,\bz)}\frac{(E-\mathcal{E}+A)^2}{2\hbar^2
|\dot{z}|^2}\right\},\end{align} where we have used
$\mathcal{B}(nT)=-2i\hbar j \ln(1+|z|^2)$. The poles of this
function are determined by the condition \be \label{ebk}
(\cS+\cI)(E_n)=(2n+1)\pi\hbar,\ee which is exactly the
Bohr-Sommerfeld quantization rule obtained recently in
\cite{prl92ag2004}. The residues at each pole give the Husimi
functions \begin{align}
\label{hus}\mathcal{H}_{n}(z)&=\frac{\sqrt{\pi}}{j}\frac{1+|z|^2}{|\dot{z}|}
\frac{1}{[T(E_n)+\left.(d\cI/d\mathcal{E})\right|_{E_n}]}\nonumber\\
&\times\exp\left\{-\frac{1}{g(z,\bz)}\frac{(E_n-\mathcal{E}(z)+A(z))^2}{2\hbar^2
|\dot{z}|^2}\right\}.\end{align} Here we have multiplied by the
coherent states normalization factor $(1+|z|^2)^{-2j}$. These
functions are our main result. They are in general not normalized,
and have a strong resemblance with the canonical semiclassical
Husimi functions presented in \cite{jpa34mb2001}.

As a first example, we calculate the Husimi function for the
simple case $\hat{H}=\hbar\omega J_z$. The Hamilton equations can
easily be solved and give $z(t)=e^{-i\omega t}z_i$ and
$\bz(t)=e^{i\omega (t-T)}\bz_f$, which implies $|\dot{z}|=\omega
|z|$. The SK correction is also very simple, with
$A(z)=\hbar\omega/2$ and $d\mathcal{I}_{SK}/d\mathcal{E}=0$. The
final result is \be \mathcal{H}_{m}(r)=\frac{1+r}{2j\sqrt{\pi r}}
\exp\{-\frac{(j(1-r)+(1+r)(m+1/2))^2}{4jr}\},\ee where $m$ goes
from $-j$ to $j$ and $r=|z|^2$. When properly normalized, this
approximates the exact distribution, \be
\mathcal{H}^{E}_{m}(r)=\frac{(2j)!}{(j+m)!(j-m)!}
\frac{r^{j+m}}{(1+r)^{2j}},\ee quite well for large values of $j$,
as we can see in Fig 1.

Now let us turn our attention to a less trivial system. In
\cite{prl92ag2004} the authors have shown that the semiclassical
quantization condition (\ref{ebk}) works very well for the
Lipkin-Meshkov-Glick (LMG) model \be \hat{H}=\hbar \omega
J_z+\alpha\hbar^2[J_x^2-J_y^2],\ee already at moderate values of
$j$. We now consider the accuracy of the semiclassical
approximation (\ref{hus}) for its stationary states. For small
values of $\alpha$ the results are very similar to the previous
case, so we consider only $j=200$, $\omega=1$ and
$\hbar\alpha=1000$. In order to display the results, we use the
canonical coordinates $(q,p)$ given in (\ref{cano}), in terms of
which the phase space is compact, $q^2+p^2\leq 4\hbar j$. We show
the exact and the semiclassical (normalized) Husimi functions for
two different states in Fig 2 and Fig 3. The agreement is
excellent.

Summarizing, we have obtained a semiclassical approximation for
the phase space representation of stationary states of spin
systems. This was done by investigating the semiclassical Green
function in the vicinity of real periodic trajectories. The
accuracy of the result was verified by comparing it with exact
calculations for the simple case $\hat{H}=\hbar\omega J_z$ and for
the Lipkin-Meshkov-Glick model. Husimi functions are known to be
good tools to study quantum chaos \cite{jpa23pf1990}, and an
extension of this theory to more degrees of freedom would be
interesting in order to approach chaotic systems (a trace formula
for chaotic spin systems was recently obtained \cite{prl89mp2002}
but, as already noted, taking the trace obliterates the
information about the residues). Work in this direction is in
progress.

We acknowledge financial support from Fapesp (Funda\c{c}\~ao de
Amparo \`{a} Pesquisa do Estado de S\~ao Paulo).

\end{document}